\documentclass[prl,aps,showpacs,nofootinbib,twocolumn,preprintnumbers]{revtex4}
\usepackage{graphicx}
\usepackage{amsfonts}
\usepackage{amssymb}
\usepackage{latexsym}
\usepackage{color}
\input{colordvi.tex}

\newcommand{\CO}{{\cal O}}

\newcommand{\bear}{\begin{array}}  \newcommand{\eear}{\end{array}}
\newcommand{\bea}{\begin{eqnarray}}  \newcommand{\eea}{\end{eqnarray}}
\newcommand{\beq}{\begin{equation}}  \newcommand{\eeq}{\end{equation}}
\newcommand{\bef}{\begin{figure}}  \newcommand{\eef}{\end{figure}}
\newcommand{\bec}{\begin{center}}  \newcommand{\eec}{\end{center}}
\newcommand{\non}{\nonumber}  
\newcommand{\lmk}{\left(}  \newcommand{\rmk}{\right)}
\newcommand{\lkk}{\left[}  \newcommand{\rkk}{\right]}

\newcommand{\del}{\partial}  

\newcommand{\bib}{\bibitem}

\def\IBB#1#2#3{{\bf #1}, #2 (20#3)}

\def\IBIDD#1#2#3{{\it ibid}. {\bf #1}, #2 (20#3)}

\def\APJSS#1#2#3{Astrophys. J. Suppl. Ser. {\bf #1}, #2 (20#3)}

\def\CQG#1#2#3{Class. Quantum Grav. {\bf #1}, #2 (19#3)}
\def\CQGG#1#2#3{Class. Quantum Grav. {\bf #1}, #2 (20#3)}

\def\JCAPP#1#2#3{J. Cosmol. Astropart. Phys. {\bf #1}, #2 (20#3)}

\def\PLB#1#2#3{Phys. Lett. B {\bf #1}, #2 (19#3)}

\def\PLBold#1#2#3{Phys. Lett. {\bf#1B}, #2 (19#3)}

\def\PRD#1#2#3{Phys. Rev. D {\bf #1}, #2 (19#3)}
\def\PRDD#1#2#3{Phys. Rev. D {\bf #1}, #2 (20#3)}

\def\PRLL#1#2#3{Phys. Rev. Lett. {\bf#1}, #2 (20#3)}

\def\PRT#1#2#3{Phys. Rep. {\bf#1}, #2 (19#3)}

\def\PTPS#1#2#3{Prog. Theor. Phys. Suppl. {\bf #1}, #2 (19#3)}

\begin{document}
\preprint{FTPI-MINN-07/21, UMN-TH-2610/07}
\title{
 D-term chaotic inflation in supergravity}

\author{Kenji Kadota}
\affiliation{William I. Fine Theoretical Physics
Institute, University of Minnesota, Minneapolis, MN 55455}

\author{Masahide Yamaguchi} 
\affiliation{Department of Physics and Mathematics, Aoyama Gakuin
University, Sagamihara 229-8558, Japan}


\begin{abstract}
Even though the chaotic inflation is one of the most popular inflation
models for its simple dynamics and compelling resolutions to the initial
condition problems, its realization in supergravity has been considered
a challenging task. We discuss how the chaotic inflation dominated by
the D-term can be induced in supergravity, which would give a new
perspective on the inflation model building in supergravity.
\end{abstract}

\pacs{98.80.Cq}

\maketitle

Cosmic inflation has been one of the most successful early universe
scenarios for more than 25 years, with the ever-growing supports from
the observations of the cosmic microwave background anisotropies and the
large scale structure of the universe \cite{inflation,WMAP3,sdss}. Among
many types of inflation models proposed so far, chaotic inflation is
special for its amelioration of the initial condition problems
\cite{chaotic} and it would be of considerable interest to realize chaotic
inflation in a sensible particle physics theory.

One of the leading theories as an extension of the minimal standard
model is supersymmetry \cite{SUSY} which gives the attractive solutions
to the hierarchy problem of the standard model as well as the
unification of three gauge couplings. In particular, in the early
universe, its local version supergravity would govern the dynamics of
the universe, while there has been criticisms for the implementation of
chaotic inflation in supergravity. This is simply because the scalar
potential coming from the F-term has an exponential dependence on the
K\"ahler potential. This prevents the scalar fields from acquiring the
amplitudes larger than the reduced Planck scale $M_{p} \simeq 2.4 \times
10^{18}$GeV and also spoils the flatness of an inflaton potential (so
called $\eta$-problem). In order to circumvent this difficulty
\cite{early}, Kawasaki, Yanagida and one of the present authors (M.Y.)
introduced the Nambu-Goldstone-like shift symmetry \cite{KYY}, so that
the imaginary part of the an inflaton field does not suffer from the
exponential growth. Though such a shift symmetry is motivated from the
string theory and is similar to, for example, the Heisenberg symmetry
\cite{heisenberg}, it may be difficult to associate it with the low
energy effective theory of particle physics such as grand unified theory
(GUT). It would be then worth seeking an alternative supergravity
chaotic inflation model without such a symmetry possibly absent in the
effective field theory.

The scalar potential in supergravity also consists of, in addition to
the F-term, the D-term which does not have an aforementioned dangerous
exponential factor. A plausible possibility to realize the supergravity
chaotic inflation then would be to consider the inflation models where
the D-term dominates over the F-term. In the conventional D-term
inflation models \cite{stewart,BD,Halyo,dterm}, the energy density is
sourced by the Fayet-Iliopoulos (FI) term $\xi$ in D-term and the slope
of an inflaton potential is induced by the one-loop corrections. Since
the one-loop corrections cannot exceed the tree level potential energy
density of order $\xi^2 \ll M_p^4$, the inflation cannot start from the
Planckian energy scale. This in turn implies that chaotic inflation is
not compatible with the standard D-term inflation models because the
universe would collapse before such an inflation energy scale is reached
unless the universe started with an open geometry. It would be then
intriguing to consistently incorporate chaotic inflation in a D-term
dominated inflation model, possibly without making use of the FI term
dominance or one-loop corrections.

In this paper, we present a chaotic inflation model in supergravity
where the D-term dominates the inflation energy density. The D-term is
helpful for the realization of chaotic inflation by allowing the
(beyond) Planckian amplitude for an inflaton field as well as for the
avoidance of the $\eta$-problem. Our D-term chaotic inflation model
requires neither the shift symmetry nor the dominance of FI term. No
need for the FI term dominance in our model alleviates the potentially
unnatural tuning of its amplitude $\xi$ which otherwise would be
constrained tightly by the cosmic perturbations and cosmic strings as in
the conventional D-term inflation models \cite{comments}.

We introduce four superfields $\Phi_{i}$ ($i = 1, 2, 3, 4$) charged
under $U(1)$ gauge symmetry and (global) $U(1)_R$ symmetry. The charges
$Q_i,Q^{R}_{i}$ of the superfields are given in Table \ref{tab:charges}
which ensure our toy model is anomaly free \cite{KK,try}. Then, the
general (renormalizable) superpotential for these fields is given by
\beq
  W = a \Phi_1^2 \Phi_3 - b \Phi_2 \Phi_3 + c \Phi_2 \Phi_4^2,
  \label{eq:super}
\eeq
where we set the constants $a, b, c$ to be real and positive for
simplicity and a non-renormalizable term $\Phi_1^2 \Phi_4^2$ is omitted
since it does not change the dynamics essentially.
\begin{table}[btc]
  \begin{center}
    \begin{tabular}{| c | c | c | c | c |}
        \hline
                   & $\Phi_1$ & $\Phi_2$ & $\Phi_3$ & $\Phi_4$ \\
        \hline
        $Q$        & $+1$     & $+2$     & $-2$     & $-1$ \\
        \hline 
        $Q^R$      & $0$      & $0$      & $+2$     & $+1$ \\
        \hline 
    \end{tabular}    
    \caption{The charges of various superfields of $U(1) \times U(1)_R$.}  
    \label{tab:charges}
  \end{center}
\end{table}
Taking the canonical K\"ahler potential, $K(\Phi_i,\Phi_i^{\ast}) = \sum_{i}
|\Phi_i|^2$, and the minimal gauge kinetic
function, $f_{ab}(\Phi_i) = \delta_{ab}$, leads to the scalar potential 
consisting of the F-term $V_F$ and D-term $V_D$
\bea
  V &=& V_F + V_D, \non \\ 
  V_F  &=&
       e^{K} \lkk \,
         \left| 2 a \phi_1 \phi_3 + \phi_1^{\ast} W \right|^2 
       + \left| - b \phi_3 + c \phi_4^2 + \phi_2^{\ast} W \right|^2
        \right. \non \\
       &+& \left.
       \left| a \phi_1 - b \phi_2 + \phi_3^{\ast} W \right|^2 
       + \left| 2 c \phi_2 \phi_4 + \phi_4^{\ast} W \right|^2 
       - 3 |W|^2 \,
            \rkk
        , \non \\
  V_D&=& \frac{g^2}{2} 
       \lmk |\phi_1|^2 + 2 |\phi_2|^2 - 2 |\phi_3|^2 - |\phi_4|^2 \rmk^2,
\eea
where $g$ is the gauge coupling of the $U(1)$ symmetry, and we take the
vanishing FI term for simplicity. Here and hereafter we set the reduced
Planck scale $M_p$ to be unity.\footnote{The qualitative features of our
chaotic inflation model would not be affected even if the
non-renormalizable terms, such as $\lambda \phi^4 (\phi/M_p)^n$, appear
in the potential as long as $\lambda \ll 1$, that is, the effective
cut-off scale is larger than the reduced Planck scale. Such a small
$\lambda$ associated with the breaking of the R-symmetry used in our toy
model to prohibit the non-renormalizable terms is natural in 't Hooft's
sense.}

The minimum of the F-term (the F-flat condition) is given by
\beq
  V_F = 0 \Longleftrightarrow 
    \left\{
      \begin{array}{c}
        a \phi_1^2 - b \phi_2 = 0 \quad\&\quad \phi_3 = \phi_4 = 0. \\
        {\rm or} \\
        \phi_1 = \phi_2 = 0 \quad\&\,\,\, - b \phi_3 + c \phi_4^2 = 0. 
      \end{array} 
    \right.
\eeq
On the other hand, the minimum of the D-term (the D-flat condition) is
given by
\beq
    V_D = 0 \Longleftrightarrow 
    |\phi_1|^2 + 2 |\phi_2|^2 = 2 |\phi_3|^2 + |\phi_4|^2.  
\eeq
Hence, the global minimum of the potential is given by
\beq
  \phi_1 = \phi_2 = \phi_3 = \phi_4 = 0.
\eeq

However, when the universe starts around the Planck scale, if $|\phi_1|
\gg 1$ or $|\phi_2| \gg 1$, and $|\phi_3|, |\phi_4| \lesssim 1$ for
example, the almost F-flat condition ($a\phi_1^2 \simeq b\phi_2,
\phi_3=\phi_4=0$) is first realized due to the exponential factor
$e^{K}$ of the F-term. Consequently, the potential is mostly dominated by the
D-term and chaotic inflation can take place.

Because the system is invariant under the following
transformation
\beq
  \phi_1 \leftrightarrow \phi_4, \quad
  \phi_2 \leftrightarrow \phi_3, \quad
  a \leftrightarrow c, \quad
  Q \rightarrow -Q,
\eeq
the dynamics is essentially the same even if we interchange
$\phi_1$ and $\phi_2$ by $\phi_3$ and $\phi_4$. Thus, we concentrate on
the case that $|\phi_1| \gg 1$ or $|\phi_2| \gg 1$, and $|\phi_3|,\,
|\phi_4| \lesssim 1$.

Now, we investigate the dynamics in details.  Despite the $e^K$ factor
of F-terms, due to the presence of the relatively rather small but
non-vanishing D-terms, the actual inflaton trajectory is slightly
deviated from the exact F-flat direction and given by solving the
equations (1) $\del V / \del \phi_{i}^{\ast} = 0$($i = 2,3,4$) or (2)
$\del V / \del \phi_{i}^{\ast} = 0$($i = 1,3,4$), depending on the
magnitude of $b/a$, as clarified later. The solution (named M1) of the
first equations is given by $\phi_2 = \phi_2(\phi_1), \phi_3=\phi_4=0$
and that (named M2) of the second equations is given by $\phi_1 =
\phi_1(\phi_2), \phi_3=\phi_4=0$.

Here, we check whether inflation indeed occurs along this field trajectory. 
For this purpose, we first evaluate the mass terms of the
fields $\phi_3$ and $\phi_4$ along these trajectories M1 and M2,
$V_{,ij}|_M \phi_{i}^{\ast} \phi_{j}$ with $V_{,ij} \equiv \del^2
V/(\del\phi_i^{\ast} \del\phi_j)$. The suffix $M$ represents the
evaluations along either the trajectory M1 or M2. Then, the
mass matrix of the fields $\phi_3$ and $\phi_4$, $V_{,ij}|_M$, is given
by
\bea
  V_{,33}|_M &\simeq& 
      e^{K} (4 a^2 |\phi_1|^2 + b^2), \quad
   V_{,3i}|_M = 0\,\,\,({\rm for}\,\,\,i = 1, 2, 4),  \non \\
  V_{,44}|_M &\simeq& 
      4 e^{K} c^2 |\phi_2|^2, \quad
   V_{,4i}|_M = 0\,\,\,({\rm for}\,\,\,i = 1, 2, 3).
  \label{eq:matrix34}
\eea
Both of these masses are much larger than $H^2 \simeq g^2
(|\phi_1|^2+2|\phi_2|^2)^2/2$ unless the constants $a,b,c$ are
exponentially small (typically $e^{-g^{-1}}, g =\CO(10^{-6})$), which makes
$\phi_3$ and $\phi_4$ quickly go to the zeros. As a result, we can safely set
$\phi_3$ and $\phi_4$ to be zero and we can discuss the dynamics of the inflaton
based on the following
potential,
\beq
  V_{\rm eff}(\phi_1,\phi_2) \equiv V(\phi_1,\phi_2,\phi_3=0,\phi_4=0).
\eeq
By use of the $U(1)$ gauge symmetry, we can, for instance, make the
field $\phi_1$ real without loss of generality, so that the ${\rm Im}
\phi_2$ rapidly goes to the zero because the effective mass squared of
the imaginary part of $\phi_2$ is given by $m^2_{{\rm Im} \phi_2} \simeq
b^2 e^{K}$. We therefore consider the following effective potential by
redefining the fields $\phi_i \equiv \sqrt{2}\, {\rm Re}\,\phi_i$ $(i =
1, 2$, and we take $\phi_i$ to be positive for definiteness) and $b'
\equiv \sqrt{2} b$, \beq V_{\rm eff}(\phi_1,\phi_2) = \frac14 e^{K} \lmk
a \phi_1^2 - b' \phi_2 \rmk^2 + \frac{g^2}{8} \lmk \phi_1^2 + 2 \phi_2^2
\rmk^2 \eeq
with $K = (\phi_1^2 + \phi_2^2) / 2$ and the canonical kinetic terms.

Now, we would like to discuss the dynamics of the inflation based on the
above potential. First of all, we consider the region where $B \equiv
b'/a \gg \phi_1 (\gg 1)$ is satisfied. As shown later, in this region,
the field $\phi_1$ plays the role of the inflaton while the inflationary
trajectory is almost determined by the condition (M1) $\del V_{\rm eff}
/ \del \phi_2 = 0$, which is equivalent to
\bea
  a\phi_1^2 - b'\phi_2 &=& 
    e^{-K} 
    \frac{g^2 (\phi_1^2+2\phi_2^2) \phi_2}
         {\frac12 b' - \frac14 \phi_2 (a\phi_1^2 - b'\phi_2)}
              \non \\ 
                         &\simeq&
    e^{-K} \frac{2}{b'} g^2 (\phi_1^2+2\phi_2^2) \phi_2.
\eea     
Here note that combining $a\phi_1^2 - b' \phi_2 = \CO(e^{-K})$ with $B =
b'/a \gg \phi_1$ leads to $\phi_1 \gg \phi_2$.  The F-term contribution
to the potential is estimated as
\bea
  V_F &=& \frac{\frac14 \phi_2 (a\phi_1^2 - b'\phi_2)}
         {\frac12 b' - \frac14 \phi_2 (a\phi_1^2 - b'\phi_2)}
        g^2 (\phi_1^2+2\phi_2^2)  \non \\
      &<& g^2 (\phi_1^2+2\phi_2^2) 
      \ll V_D = \frac{g^2}{2} (\phi_1^2+2\phi_2^2)^2
\eea
for $\phi_1 \gg 1$. Thus, the potential is dominated by the D-term
during inflation. We also consider the mass matrix of the fields
$\phi_1$ and $\phi_2$, $V_{,ij}|_{M_1}$($V_{,ij} \equiv \del^2 V_{\rm
eff}/(\del\phi_i\del\phi_j)$), which is given by
\bea
  V_{,11}|_{M1} &\simeq& 2a^2\phi_1^2 e^{K} 
    + 2 \frac{a}{b'} g^2 (\phi_1^2+2\phi_2^2) \phi_2 (1+2\phi_1^2)
       \non \\
    && + \frac{g^2}{2} (3\phi_1^2+2\phi_2^2), \non \\ 
  V_{,12}|_{M1} &\simeq& -ab' \phi_1 e^{K} 
    + 2 \frac{a}{b'} g^2(\phi_1^2+2\phi_2^2) \phi_1\phi_2 (\phi_2-\frac{b'}{2a})
       \non \\    
    && + 2 g^2 \phi_1\phi_2, \non \\ 
  V_{,22}|_{M1} &\simeq& \frac12 b'^2 e^{K} 
    - 2 g^2 (\phi_1^2+2\phi_2^2) \phi_2^2
    + g^2 (\phi_1^2+6\phi_2^2)
  \label{eq:matrix12}
\eea
up to the order of $\CO((e^{K})^0)$.\footnote{More precisely, we have
used the approximation that $b' \gg |\phi_2(a\phi_1^2-b'\phi_2)|$ and
$a, b'\phi_2 \gg |a\phi_1^2-b'\phi_2|$.} The effective mass squared
$\lambda$ of the fields $\phi_1$ and $\phi_2$ is given as the solutions
of the following equation,
\beq
  \lambda^2 - (V_{,11} + V_{,22}) \lambda 
    + V_{,11} V_{,22} - V_{,12}^2 = 0,
\eeq
where
\bea
  V_{,11} + V_{,22}|_{M1} &\simeq& 
   e^{K} \lmk 2a^2\phi_1^2 + \frac12 b'^2 \rmk \simeq \frac12 b'^2 e^{K},
                      \non \\
  V_{,11} V_{,22} - V_{,12}^2|_{M1} &\simeq& 
    \frac34 g^2 b'^2 \phi_1^2 e^{K}                        
  \label{eq:lambda}
\eea     
up to the order of $\CO(e^{K})$. Here, we have used the approximation
that $a\phi_1^2 - b' \phi_2 = \CO(e^{-K})$, $B = b'/a \gg \phi_1$, and
$\phi_1 \gg \phi_2$. The effective squared masses are then approximately given
by
\beq
  \lambda \simeq \frac12 b'^2 e^{K}, \qquad \frac32 g^2 \phi_1^2 \ll H^2
  \simeq V_D / 3,
\eeq
where $H$ is a Hubble parameter. The inflaton field in the chaotic
inflation corresponds to this effectively massless mode. This light mass
squared vanishes for $g=0$ as expected, reflecting the exact F-flat
direction.

Since $V_{,22} \gg V_{,11}$ and $\phi_1 \gg \phi_2$, the inflationary
trajectory is given by the minimum of the field $\phi_2$, $\del V / \del
\phi_2 = 0$, which enables us to write the minimum of $\phi_2$ as a
function of $\phi_1$, $\phi_2^m = \phi_2^m(\phi_1)$.  The field
trajectory governing the inflation dynamics therefore can be
parameterized by the field $\phi_1$ which we call an
inflaton.\footnote{Note here that the effectively massless field
trajectory parameterized by the inflaton field $\phi_1$ is different
from the $\phi_1$ direction with the mass $V_{,11} \gg H^2$.} Then, by
inserting the above relation into the effective potential, we define the
reduced potential $V_{\rm r1}(\phi_1)$ as
\beq
  V_{\rm r1}(\phi_1) \equiv V_{\rm eff}(\phi_1,\phi_2^m(\phi_1))
                     \lmk \simeq \frac{g^2}{8} \phi_1^4 \rmk.  
\eeq
As explicitly shown in Ref. \cite{YY}, when 
there is only one massless mode and the
other modes are massive, the generation of adiabatic density fluctuations as
wells as the dynamics of the homogeneous mode is completely determined
by the reduced potential $V_{\rm r1}(\phi_1)$. Indeed, the equation of
motion for the homogeneous mode of the inflaton $\phi_1$ along the
rolling direction (M1) is approximated as
\beq
  \left.
    \ddot{\phi}_1 + 3 H \dot{\phi}_1 + \frac{\del V_{\rm eff}}{\del \phi_1}
  \right|_{M1}
  =
    \ddot{\phi}_1 + 3 H \dot{\phi}_1 
      + \frac{dV_{\rm r1}}{d\phi_1} 
  = 0,
 \label{eq:homo}
\eeq
where the dot represents time derivative. Thus, the dynamics of
the inflation with the inflaton $\phi_1$ can be estimated by using the
reduced potential $V_{\rm r1}(\phi_1)$ as long as the dynamics rolls
down along the minimum of $\phi_2$ (M1).

Next, we evaluate the primordial density fluctuations in the
longitudinal gauge. The equation of motion for the perturbation
$\delta\phi_i$ of each real field is given by \cite{fluc}
%
\bea
  && \ddot{\delta\phi}_i + 3 H \dot{\delta\phi}_i 
       - \frac{\nabla^2}{a^2} \delta\phi_i + 
        \sum_j \left.
          \frac{\del^2 V_{\rm eff}}{\del\phi_j \del\phi_i} 
                    \right|_{M1} \delta\phi_j \non \\
  &&
   \qquad \qquad 
      \left.
      = -2 \frac{\del V_{\rm eff}}{\del\phi_i} 
                    \right|_{M1} \Phi + 4 \dot{\phi}_i\dot{\Phi}, 
  \label{eq:multifluc}
\eea
%
where $\Phi$ is the gravitational potential. We hereafter use
the same symbol $\phi_i$ for both the homogeneous mode and the full field
for notational brevity unless stated otherwise.

We are interested only in the adiabatic density fluctuations characterized by the
condition
\beq
  \frac{\delta\phi_1}{\dot{\phi}_1} = \frac{\delta\phi_2}{\dot{\phi}_2}
  \quad \Longleftrightarrow \quad
  \delta\phi_2 = \frac{d\phi_2^m(\phi_1)}{d\phi_1} \delta\phi_1
  \label{eq:ad}
\eeq
where we have used
$\dot{\phi}_2/\dot{\phi}_1=d\phi_2^m(\phi_1)/d\phi_1$. Since the
relation $\del V_{\rm eff}(\phi_1,\phi_2^m(\phi_1)) / \del \phi_2 = 0$
holds for any $\phi_1$ in the relevant region, we find
\beq
  \left.
  \frac{d}{d\phi_1} \lkk 
      \frac{\del V_{\rm eff}}{\del \phi_2}(\phi_1,\phi_2^m(\phi_1))
                    \rkk
  = V_{,12} + V_{,22} \frac{d\phi_2^m}{d\phi_1} \right|_{M1}
  = 0.
  \label{eq:relation} 
\eeq
%
%
Taking into account this relation and
\bea
  \frac{d^2V_{\rm r1}}{d\phi_1^2}
   &=& \left. V_{,11} + 2 \frac{d\phi_2^m}{d\phi_1} V_{,12}
     + \lmk \frac{d\phi_2^m}{d\phi_1} \rmk^2 V_{,22} \right|_{M1} 
     \non \\
   &=& \left. \frac{V_{,11}V_{,22}-V_{,12}^2}{V_{,22}} \right|_{M1}
   \lmk \simeq \frac32 g^2 \phi_1^2 \rmk,
\eea
the equation of motion for the perturbation $\delta\phi_1$ becomes
\bea
  \ddot{\delta\phi}_1 + 3 H \dot{\delta\phi}_1 
       - \frac{\nabla^2}{a^2} \delta\phi_1 + 
        \frac{d^2V_{\rm r1}}{d\phi_1^2} \delta\phi_1
      = -2 \frac{dV_{\rm r1}}{d\phi_1} \Phi + 4 \dot{\phi}_i\dot{\Phi}. 
  \label{eq:fluc2}
\eea
Thus, the perturbation $\delta\phi_1$ is completely determined by the
reduced potential $V_{\rm r1}(\phi_1)$. Note that $d^2V_{\rm
r1}/d\phi_1^2$ is the effective mass squared along the rolling direction
given by $\del V_{\rm eff} / \del \phi_2 = 0$. Therefore, this rolling
direction actually coincides with the eigenvector of the effectively
massless mode of $\lambda$.

On the other hand, by use of the adiabatic condition, the gravitational
potential is described only by $\delta\phi_1$ as
\bea
  \lmk \dot{H} - \frac{\nabla^2}{a^2} \rmk \Phi =
    \frac{1}{2M_G^2} 
       \lmk \ddot{\phi}_1 \delta\phi_1 - \dot{\phi}_1 \dot{\delta\phi}_1 \rmk 
       \lkk 1 + \lmk \frac{d\phi_2^m}{d\phi_1} \rmk^2 \rkk.
 \label{eq:gpdirect} 
\eea
In the long wave limit, $\Phi \simeq (d/dt)(\delta\phi_1/\dot{\phi}_1)$,
where we have used $\dot{H} = - (\dot{\phi}_1^2+\dot{\phi}_2^2) /
(2M_G^2) = - \dot{\phi}_1^2 [1 + (d\phi_2^m/d\phi_1)^2] / (2M_G^2)$.
%
%
This expression of the gravitational potential coincides with
that of the single field inflation with the reduced potential $V_{\rm
r1}(\phi_1)$. We can in consequence calculate the density fluctuations of our
inflation model
based on the reduced potential $V_{\rm r1}(\phi_1) \simeq g^2
\phi_1^4 / 8$.

In the opposite region where $B = b'/a \ll \phi_1 (\ll \phi_2)$, the
inflationary trajectory is given by the minimum of $\phi_1$, $\phi_1^m =
\phi_1^m(\phi_2)$ coming from the condition (M2) $\del
V_{\rm eff}/\del\phi_1 = 0$. The field $\phi_2$ now plays a role of
the inflaton. Defining the reduced potential $V_{\rm r2}(\phi_2) \equiv
V_{\rm eff}(\phi_1^m(\phi_2),\phi_2)$, we can easily show that
the generation of adiabatic density fluctuations as wells as the dynamics of
the homogeneous mode can be calculated by using the reduced potential
$V_{\rm r2}(\phi_2)$.

For $B > 1$, in the region where $B \sim \phi_1$, the use of the
condition $\del V_{\rm eff}/\del\phi_1 = 0$ or $\del V_{\rm
eff}/\del\phi_2 = 0$ may generate the error of the effective mass
squared up to the factor 2 due to $V_{,11} \sim V_{,22}$.
However, since the true trajectory is effectively massless, we can see 
from the mass matrix that the deviation is at most
${\cal O}(e^{-K})$. It therefore still gives an approximate trajectory of the
inflaton, depending on $\phi_1 > \phi_2$ or $\phi_1 < \phi_2$ and hence
the qualitative behaviors of the inflaton dynamics and the density
fluctuations do not change in this region. Hence the chaotic inflation can be induced in a wide range of the parameters in our toy model. It could also indicate that, once a model possesses a F-flat direction lifted by a D-term, the chaotic inflation could be induced without so much restrictions on the model parameters besides those from the cosmic perturbations.

Finally we give a few comments on our model. The standard procedure
shows that the gauge coupling $g$ should be $g \sim 10^{-6}$ in order to
explain the primordial density fluctuations. This value of the gauge
coupling is much smaller than the standard gauge couplings. However,
this may not be a problem because the gauge symmetry may be a hidden
gauge symmetry, or the gauge coupling could be suppressed, for instance,
by considering the extra dimensions.  Next, even though we presented a
toy model of the quartic potential chaotic inflation, the leading order
polynomial can be different by an appropriate choice of the non-minimal
gauge kinetic function (for instance, a form $f = 1 + d_i |\phi_i|^2$
($d_i$ : const) could lead to a quadratic potential). Note this is in
contrast to the standard D-term inflation with the dominance of the FI
term where a non-minimal gauge kinetic function spoils the flatness of
the inflaton potential \cite{mth}.  As for the reheating, it may require
the spontaneous breaking of the gauge symmetry after inflation because
the inflaton cannot directly decay into the standard particles due to
the charge conservation.  Such a breaking can occur, for instance, by
the introduction of the FI term or a Higgs-like field, and such
modifications of our simple toy model presented in this paper would lead
to a variety of inflation models with the potentially rich
phenomena.\footnote{For instance, an inflaton superpotential of form $W
= (a \Phi_1^2 - b \Phi_2^2) \Phi_3$ or $W = M (a \Phi_1 - b \Phi_2)
\Phi_3$, with a non-vanishing FI term, could lead to a simpler chaotic
inflation dynamics with a subsequent hybrid inflation stage, even though
one needs to introduce the additional fields to cancel anomaly.} Further
study will be given in the forthcoming paper \cite{full}.

In summary, we have shown that chaotic inflation can take place in
supergravity even without the shift symmetry where the inflaton field
trajectory follows the (almost) F-flat direction lifted by the D-term.
We stress that, according to our analysis demonstrated through a toy model, many of such F-flat directions could potentially cause the D-term chaotic inflation.
The D-term is related to the gauge coupling and a variant of our model
could lead to a possible link between a chaotic inflation model and the
low energy effective theory of particle physics \cite{DGUT}.


We thank J. Giedt, M. Kawasaki, H. Murayama, E. Stewart, F. Takahashi,
and J. Yokoyama for the useful discussions. K.K. is supported by DOE
grant DE-FG02-94ER-40823 and M.Y. is supported in part by JSPS
Grant-in-Aid for Scientific Research No.~18740157 and by the project of
the Research Institute of Aoyama Gakuin University.

\end{document}